\documentclass[aps,prb,amsmath,amssymb,showpacs,reprint,preprintnumbers,nobalancelastpage]{revtex4-1}
\usepackage{hyperref}
\hypersetup{
	pdfpagemode=UseNone,
	pdfstartview=FitH
}

\DeclareMathOperator{\qdim}{\text{dim}_q}
\newcommand{\Integer}{\mathbb{Z}}
\newcommand{\Complex}{\mathbb{C}}
\newcommand{\mat}{\begin{pmatrix}}
\newcommand{\tam}{\end{pmatrix}}
\DeclareMathOperator{\tr}{{\text{tr}}}
\newcommand{\bS}{\mathbb{S}}
\newcommand{\bI}{\mathbb{I}}
\newcommand{\qsu}{\cU_q\bigl[su(2)\bigr]}
\newcommand{\qg}{\cU_q[\mathfrak{g}]}
\newcommand{\idop}{\mathbb{I}}

\newcommand{\cU}{\mathcal{U}}
\newcommand{\cV}{\mathcal{V}}
\newcommand{\g}{\mathfrak{g}}

\newcommand{\qAKLT}{$q\text{AKLT}$ }

\begin{document}

\title{Symmetry protected topological phases beyond groups:\\The $q$-deformed Affleck-Kennedy-Lieb-Tasaki model}

\author{Thomas Quella}%
\email{Thomas.Quella@unimelb.edu.au}
\affiliation{ The University of Melbourne, School of Mathematics and Statistics, Parkville 3010 VIC, Australia }
 
\date{\today}

\begin{abstract}
  We argue that the $q$-deformed spin-$1$ AKLT Hamiltonian should be regarded as a representative of a symmetry-protected topological phase. Even though it fails to exhibit any of the standard symmetries known to protect the Haldane phase it still displays all characteristics of this phase: Fractionalized spin-$\frac{1}{2}$ boundary spins, non-trivial string order and -- when using an appropriate definition -- a two-fold degeneracy in the entanglement spectrum. We trace these properties back to the existence of an $SO_q(3)$ quantum group symmetry and speculate about potential links to discrete duality symmetries. We expect our findings and methods to be relevant for the identification, characterization and classification of other symmetry-protected topological phases with non-standard symmetries.
\end{abstract}


\preprint{\eprint{arXiv:2005.09072}}

\maketitle

\section{Introduction}

  The Haldane phase of antiferromagnetic spin-$1$ $SU(2)$ quantum spin chains is one of the prototypes of symmetry-protected topological (SPT) phases.\cite{Pollmann:2012PhRvB..85g5125P,Schuch:1010.3732v3,Chen:PhysRevB.84.235128} It exhibits a unique ground state, an excitation gap, a two-fold degeneracy in the bipartite entanglement spectrum and, at least when $SO(3)$ spin-rotation symmetry is preserved, fractional boundary spins as well as non-trivial string order. While these facts have only been verified numerically for the spin-$1$ Heisenberg model, the associated AKLT model \cite{Affleck:PhysRevLett.59.799,Affleck:1987cy} provides an alternative representative of the same phase in which all these properties can be established with full analytical rigor and where the existence of non-trivial string order can be linked to the breaking of a hidden $\Integer_2\times\Integer_2$ symmetry.\cite{Affleck:PhysRevLett.59.799,Affleck:1987cy,DenNijs:PhysRevB.40.4709,Kennedy:10.1007/BF02097239,Pollmann:PhysRevB.81.064439} This is due to the fact that the ground state of the AKLT Hamiltonian has a simple representation as a matrix product state (MPS).\cite{Affleck:PhysRevLett.59.799,Affleck:1987cy,Fannes:1989ns,Fannes:1990px} (See also Refs.~\onlinecite{Fannes:1990ur,Perez-Garcia:2007:MPS:2011832.2011833,Schollwock:2011AnPhy.326...96S} for a general discussion of MPS and finitely correlated states.)

  In view of the simple construction and intriguing properties of the AKLT model it is no surprise that variations of this construction have been applied to other symmetries, notably quantum group deformations of the $SU(2)$ spin-rotation symmetry.\cite{Batchelor:1990JPhA...23L.141B,Klumper:1991JPhA...24L.955K,Klumper:1992ZPhyB..87..281K,Batchelor:1994IJMPB...8.3645B,Totsuka:1994JPhA...27.6443T,Fannes:10.1007/BF02101525} We will call the resulting spin-$1$ model which is based on a $\qsu$ quantum group symmetry the \qAKLT model. From the construction of the \qAKLT Hamiltonian as the parent Hamiltonian of an MPS it is evident that the model exhibits fractionalized boundary spins. Moreover, it has been known for a very long time that the model has non-trivial string order which is linked to the breaking of a generalized duality-type $\Integer_2\times\Integer_2$-symmetry.\cite{Totsuka:1994JPhA...27.6443T} Its entanglement properties \cite{Santos:2012EL.....9837005S} and correlation functions\cite{Klumper:1992ZPhyB..87..281K} have also been studied in great detail.  

  However, while these properties are all well known, at least among experts, no one so far has established a link between the \qAKLT model and SPT phases.\footnote{A connection between quantum group symmetries and SPT phases was already anticipated in Ref.~\onlinecite{Duivenvoorden:2012arXiv1206.2462D}. Speculations about the \qAKLT model (and its generalizations to higher spin) realizing an SPT phase have also appeared in Ref.~\onlinecite{Dittrich:2013arXiv1311.1798D}. However, the latter suggestion seems to boil down to a pure analogy.} As we will discuss, this is (presumably) related to the extensive breaking of standard group symmetries and, in addition, to the puzzling lack of degeneracies in its entanglement spectrum\cite{Santos:2012EL.....9837005S} that would be expected in the presence of boundary spins and quantum group symmetry. With this article we aim to close this gap. We explain why and in which sense the quantum group symmetry $\qsu$ is capable of protecting the non-trivial topology of the \qAKLT model and we revisit the entanglement calculations, showing that the existing literature has been more concerned about implementing periodic boundary conditions than about actually enforcing the quantum group symmetry. We expect that our findings will trigger a systematic search for other SPT phases with non-standard symmetries and that our methods will be helpful in identifying, characterizing and classifying them.


\section{The \qAKLT Hamiltonian and its symmetries}

  The \qAKLT model is defined on the Hilbert space of spin-$1$ quantum spins. It may be expressed in terms of standard $SU(2)$-spins $\vec{S}_i$ and in that formulation the \qAKLT Hamiltonian reads \cite{Batchelor:1990JPhA...23L.141B,Klumper:1992ZPhyB..87..281K}
\begin{align}\nonumber 
	\mathcal{H} &= b\,{\sum_{j}}\, \Bigl\{  c\,\vec{S}_j\cdot\vec{S}_{j+1} + \bigl[\vec{S}_j\cdot\vec{S}_{j+1} \nonumber \\[-3mm]
	&\qquad\quad + \tfrac{1}{2}(1-c)(q+q^{-1}-2)S_j^zS_{j+1}^z \nonumber \\ 
	&\qquad\quad + \tfrac{1}{4}(1+c)(q-q^{-1})(S_{j+1}^z-S_{j}^z)\bigr]^2 \nonumber \\ 
	&\quad + \tfrac{1}{4}\,c\,(1-c)(q+q^{-1}-2)^2(S_j^zS_{j+1}^z)^2 \nonumber \\ 
	&\quad + \tfrac{1}{4}\,c\,(1+c)(q-q^{-1})(q+q^{-1}-2)S_j^zS_{j+1}^z \nonumber \\
	&\qquad\quad \times  (S_{j+1}^z-S_j^z) \nonumber \\
	&\quad +  \tfrac{1}{4}(c-3)\bigl[\bigl(c-1+\tfrac{1}{2}(1+c)^2\bigr)S_j^zS_{j+1}^z  \nonumber \\ 
	&\qquad\quad + 2 \bigl( c- \tfrac{1}{8}(1+c)^2\bigr)\bigl((S_{j+1}^z)^2+(S_j^z)^2\bigr)\bigr] \nonumber \\
	&\quad + (c-1) + \tfrac{1}{2}\,c\,(q^{2}-q^{-2})(S_{j+1}^z-S_j^z)\Bigr\},
  \label{eq:qAKLTHamiltonian}
\end{align}
  with $c= 1 + q^{2} +q^{-2}$ and $b=[c\,(c-1)]^{-1}$. Except stated otherwise we will consider open boundary conditions or an infinite chain. The Hamiltonian is hermitian for real values of $q$. To keep the exposition simple, $q>0$ will be assumed throughout this article. We note that both parameters $b$ and $c$ are invariant under the substitution $q\to q^{-1}$.

  The standard AKLT model (obtained by setting $q=1$) has a certain number of symmetries that are known to protect its topological properties.\cite{Pollmann:2012PhRvB..85g5125P} These symmetries are i) $SO(3)$ spin-rotation symmetry,\footnote{We could say $SU(2)$ here since $SO(3)=SU(2)/\Integer_2$ and the $\Integer_2$ subgroup acts trivially on integer spins. However, from a more general perspective it is more appropriate to think of $SO(3)$ as the protecting symmetry\cite{Duivenvoorden:2012arXiv1206.2462D}} ii) its $\Integer_2\times\Integer_2$ subgroup of $\pi$-rotations around the principal axes, iii) inversion of the chain and iv) time-reversal symmetry, an antilinear symmetry implementing the transformation $\vec{S}\mapsto-\vec{S}$.

  In contrast, the Hamiltonian~\eqref{eq:qAKLTHamiltonian} is anisotropic for $q\neq1$ and hence breaks $SU(2)$ spin-rotation symmetry. The only continuous symmetry that is left is a $U(1)$-symmetry by rotations around the $z$-axis.
  The anisotropy also breaks the $\Integer_2\times\Integer_2$ symmetry group of $\pi$-rotations. Finally, inversion symmetry and time-reversal symmetry are broken by the term $(S_j^z)^2S_{j+1}^z-S_j^z(S_{j+1}^z)^2$.
  To summarize, all the discrete symmetries known to protect the Haldane phase are broken explicitly. According to the general classification,\cite{Pollmann:2012PhRvB..85g5125P,Schuch:1010.3732v3,Chen:PhysRevB.84.235128} the Hamiltonian~\eqref{eq:qAKLTHamiltonian} should thus not be regarded as residing in an SPT phase.
  
  As will be discussed in the following section, the Hamiltonian~\eqref{eq:qAKLTHamiltonian} is the parent Hamiltonian of an MPS that is constructed using the representation theory of $\qsu$. As such, it is naturally invariant under the action of this quantum group that defines a $q$-deformation of the $su(2)$ spin algebra.\footnote{Our conventions are based on what is called $\breve{\cU}_q(sl_2)$ in Ref.~\onlinecite{Klimyk:MR1492989}. We note that Ref.~\onlinecite{Klimyk:MR1492989} has reserved the symbol $\cU_q(sl_2)$ to denote the same quantum group with a different coproduct. It is worth pointing out that only the choice of coproduct used in this article is consistent with the usual hermiticity properties of spins.} This deformation is defined in terms of $q$-spin generators $\vec{\bS}$ satisfying the relations
\begin{align}
  \label{eq:CommRels}
	[\bS^z,\bS^\pm]
	=\pm\bS^\pm
	\quad\text{ and }\quad
	[\bS^+,\bS^-]
  =\frac{q^{2\bS^z}-q^{-2\bS^z}}{q-q^{-1}}\;.
\end{align}
  For the spin-$1$ representation these commutation relations are satisfied with the identification
\begin{align}
  \bS^z=S^z
  \quad\text{ and }\quad
  \bS^\pm
  =\sqrt{\frac{q+q^{-1}}{2}}S^\pm\;.
\end{align}
  Even though these expressions are just rescaled versions of the standard spin operators, there is another important signature of the $q$-deformation: The action of the $q$-spins $\vec{\bS}$ on tensor products is not simply given by adding up the $q$-spins for individual factors but rather by applying a so-called coproduct. For two tensor factors this coproduct reads
\begin{align}
	\label{eq:Coproduct}
	\Delta(\bS^z)
	&=\bS^z\otimes\bI+\bI\otimes\bS^z\nonumber\\
	\Delta(\bS^\pm)
	&=\bS^\pm\otimes q^{\bS^z}+q^{-\bS^z}\otimes\bS^\pm
\end{align}
  and the action on multiple tensor factors (such as the whole spin chain) is obtained by iterating this action appropriately. We note that the standard action of $su(2)$ is only recovered in the limit $q\to1$. A summary of useful information about $\qsu$ and its representations can be found in Appendices~\ref{ap:UqSU2} and~\ref{ap:UqSUreps}. The main result of this article is that the quantum group symmetry just sketched (or rather an incarnation adapted to the fact that we are dealing with a spin-$1$ chain) is capable of protecting the topological properties of the \qAKLT state. 
  
  It needs to be emphasized though that this symmetry is only present when the chain is (semi)infinite or considered with open boundary conditions. If periodic boundary conditions are used, as is the case in most of the literature on this subject, the Hamiltonian and also its ground state are {\em not} invariant under $\qsu$. This symmetry is only restored if specific twisted boundary conditions are used. These facts are well known in the community working on quantum group invariant integrable models, see~e.g.~Refs.~\onlinecite{Pasquier:1990NuPhB.330..523P,Karowski:1994NuPhB.419..567K,Grosse:1994JPhA...27.4761G,Links:1999JMP....40..726L}, and will also play an important role for establishing our main result. However, to keep the present discussion focused we will restrict our attention to open boundary conditions and infinite systems.

  We would also like to stress that the breaking of some of the original discrete symmetries when deforming away from $q=1$ is rather mild. The $\pi$-rotations around the $x$- and $y$-axis as well as inversion and time-reversal still leave the Hamiltonian~\eqref{eq:qAKLTHamiltonian} invariant if they are accompanied by the transformation $q\to q^{-1}$. However, these operations are not symmetries, strictly speaking, but rather duality transformations, mapping one model to another one that is physically equivalent. Such duality symmetries are not covered by the standard classification of SPT phases \cite{Pollmann:2012PhRvB..85g5125P,Schuch:1010.3732v3,Chen:PhysRevB.84.235128} and hence require a new approach.\footnote{We note that the combination of two of these tansformations is still a symmetry of the Hamiltonian~\eqref{eq:qAKLTHamiltonian}. However, inversion symmetry for instance can easily be broken by staggering the couplings without changing the ground state or its topological properties.} We defer this interesting question to future research and content ourselves with a detailed discussion of the continuous $\qsu$ symmetry. Judging from the special case $q=1$ we expect that continuous and discrete duality-type symmetries lead to the same $\Integer_2$-classification of SPT phases in the present situation.

\section{\label{sc:qAKLTState}The \qAKLT state}

  The starting point of the \qAKLT construction is a state (or rather a set of four states) that can be represented in the form of an MPS. More precisely we define
\begin{align}
	\label{eq:qAKLTstate}
	|\text{\qAKLT}\rangle_{\alpha\beta}
	=(B_1B_2\cdots B_L)_{\alpha\beta}
\end{align}
  for a finite chain of length $L$ with open boundary conditions. In this expression, $\alpha,\beta=\pm1/2$ denote the degrees of freedom associated with a left and right spin-$\frac{1}{2}$ boundary spin and the rest of the product is a mixed matrix/tensor product of matrices $B_i$ that all have the form~\cite{Klumper:1992ZPhyB..87..281K} (see Appendix~\ref{ap:MPS})
\begin{align}
  \label{eq:MPStensorB}
  B_i=\sqrt{\Omega}\mat-\frac{q^{-1}}{\sqrt{q+q^{-1}}}\,|0\rangle_i&q^{\frac{1}{2}}\,|+\rangle_i\\
  -q^{-\frac{1}{2}}\,|-\rangle_i&\frac{q}{\sqrt{q+q^{-1}}}\,|0\rangle_i\tam\;,
\end{align}
  where $\Omega=(q+q^{-1})/(1+q^2+q^{-2})$. The index $i$ indicates on which site the physical states $|0\rangle$ and $|\pm\rangle$ live. The \qAKLT state~\eqref{eq:qAKLTstate} arises from a valence bond construction involving the spin-$1$ representation as the physical spin and two spin-$\frac{1}{2}$ spins as auxiliary spins \cite{Klumper:1992ZPhyB..87..281K} (see Appendix~\ref{ap:MPS}). The associated transfer matrix has a non-degenerate (in modulus) eigenvalue~$1$ and this translates into the existence of a mass gap, see e.g.~\cite{Fannes:1990ur}.

  For completeness we note that the \qAKLT state should be defined by
\begin{align}
  \label{eq:qAKLTstatePBC}
  |\text{\qAKLT}\rangle
  =\tr\bigl(q^{2\bS^z}B_1B_2\cdots B_L\bigr)
\end{align}
  when considering closed boundary conditions. This is explained in Appendix~\ref{ap:PBC} based on equivariance properties of the MPS tensor $B$ that are derived in Appendix~\ref{ap:Equivariance}. The insertion of the twist $q^{2\bS^z}$ acting on the auxiliary space guarantees invariance under the quantum group $\qsu$ in this setting. As will be discussed in the following section it is precisely this twist (and other closely related ones) that enable us to derive entanglement properties that support the interpretation of the \qAKLT model as a representative of an SPT phase. We expect that closing the chain requires non-local terms in the Hamiltonian~\eqref{eq:qAKLTHamiltonian} if the $\qsu$-symmetry is meant to be preserved. At least this is known to be the case in integrable quantum spin chains with quantum group symmetry \cite{Karowski:1994NuPhB.419..567K,Grosse:1994JPhA...27.4761G}.

\section{\label{sc:Entanglement}The entanglement spectrum revisited}

  The entanglement properties of the \qAKLT state were discussed in Ref.~\onlinecite{Santos:2012EL.....9837005S}. These considerations were based on an MPS with periodic boundary conditions which, as we have pointed out, is not invariant under the action of $\qsu$. It is thus no surprise that the results do not reflect the degeneracies appropriately that would be expected as a result of $\qsu$-invariance and the presence of fractionalized spin-$\frac{1}{2}$ boundary spins.
  
  To avoid finite-size effects we work in an infinite chain where the MPS can be interpreted as a translation invariant iMPS.\cite{Vidal:2007PhRvL..98g0201V,Orus:PhysRevB.78.155117,Schollwock:2011AnPhy.326...96S} The MPS tensor~\eqref{eq:MPStensorB} is in right canonical form and normalized, i.e.\ it satisfies $BB^\dag=\idop$ where the product is taken both over the physical and the auxiliary index.\cite{Schollwock:2011AnPhy.326...96S} It may be checked that the alternative MPS tensor $A=\Lambda B\Lambda^{-1}$ is in left-canonical form, i.e.\ it satisfies $A^\dag A=\idop$,  where we introduced $\Lambda=\text{diag}(q^{-\frac{1}{2}},q^{\frac{1}{2}})/\sqrt{q+q^{-1}}$. The importance of these tensors lies in the fact that the two sets of semi-infinite states
\begin{align}
  |\alpha\rangle_L
  &=(\cdots A_{-3}A_{-2}A_{-1})_{\bullet\alpha}\\
  |\alpha\rangle_R
  &=(B_0B_1B_2\cdots)_{\alpha\bullet}
\end{align}
  with $\alpha=\pm1/2$
  are orthonormal on the left and right semi-infinite Hilbert space, respectively. For this reason
  the expression
\begin{align}
  \label{eq:SchmidtDecomposition}
  |\text{\qAKLT}\rangle_\infty
  &=\sum_{\alpha=\pm1/2}\Lambda_\alpha|\alpha\rangle_L\otimes|\alpha\rangle_R
\end{align}
  is a Schmidt decomposition and permits to read off the (non-degenerate) entanglement spectrum $\epsilon_\alpha=-\log\Lambda_\alpha^2$ and entanglement entropy $S_{EE}=-\sum_\alpha\Lambda_\alpha^2\log\Lambda_\alpha^2$ from the diagonal matrix $\Lambda$. We note in passing that this result precisely corresponds to the entanglement present in the normalized $\qsu$-singlet
\begin{align}
	|\text{singlet}\rangle
	=\bigl(q+q^{-1}\bigr)^{-\frac{1}{2}}\Bigl(q^{\frac{1}{2}}|\!\!\uparrow\downarrow\rangle-q^{-\frac{1}{2}}|\!\!\downarrow\uparrow\rangle\Bigr)\;.
\end{align}
  This could of course be expected since the latter is precisely what is used to describe the singlet bonds in the valence bond construction of the MPS. It is clear that this state has lower entanglement than a Bell state, at least for generic values of $q$. However, the two spins are still fully entangled in the sense that they form a singlet with respect to the quantum group symmetry, so the states are always entangled, regardless of the value of $q$, and all coefficients are completely fixed up to normalization.
  
  This type of entanglement can be captured by a $q$-deformed definition of the reduced density matrix. Following Ref.~\onlinecite{Couvreur:2017PhRvL.119d0601C} we define
\begin{align}
  \rho_R^{(q)}
  =\tr_L\bigl(q^{-2\bS_L^z}\rho\bigr)\;,
\end{align}
  where $\rho$ is the density matrix associated with $|\text{\qAKLT}\rangle_\infty$ and $\bS_L^z$ corresponds to the action of $\bS^z$ on the left part of the chain which is traced out. The MPS tensor satisfies the equivariance property
\begin{align}
  \label{eq:Equivariance}
  q^{-2\bS^z}\!\triangleright\!B
  =q^{-2\bS^z}B q^{2\bS^z}\;,
\end{align}
  where the symbol $\triangleright$ on the left hand side denotes an action on the physical space and the conjugation on the right hand side acts on the virtual spins by means of $\vec{\bS}=\vec{S}$ where $\vec{S}$ are the standard spin operators in the spin-$\frac{1}{2}$ representation. Using the (trivial) coproduct for $\bS^z$ it can then be shown that
\begin{align}
  q^{-2\bS_L^z}|\text{\qAKLT}\rangle_\infty
  &=\sum_{\alpha=\pm1/2}\Lambda_\alpha q^{2\alpha}|\alpha\rangle_L\otimes|\alpha\rangle_R\;,
\end{align}
  i.e.\ the action of $q^{-2\bS^z}$ can be pushed to the auxiliary level.
  The $q$-deformed entanglement spectrum thus reads
\begin{align}
  \epsilon_\pm^{(q)}
  =-\log\bigl[\Lambda_{\pm\frac{1}{2}}^2q^{\pm1}\bigr]
  =\log(q+q^{-1})
\end{align}
  and shows a 2-fold degeneracy which arises from the presence of virtual fractionalized spin-$\frac{1}{2}$ boundary spins.
  
  Given the $q$-deformed reduced density matrix it is then straightforward, using again the equivariance property~\eqref{eq:Equivariance}, to calculate the associated $q$-deformed entanglement entropy \cite{Couvreur:2017PhRvL.119d0601C} which is given by
\begin{align}
  S_{EE}^{(q)}
  =-\tr_R\big(q^{2\bS_R^z}\rho_R^{(q)}\log\rho_R^{(q)}\bigr)
  =\log(q+q^{-1})\;.
\end{align}
  We note that the result is just the logarithm of the so-called qantum dimension of the spin-$\frac{1}{2}$ representation describing the virtual fractionalized boundary spin. As will be shown in the next section, the full degeneracy of the $q$-deformed entanglement spectrum and the reduction of the $q$-deformed entanglement entropy to the logarithm of the quantum dimension are easily verified to generalize to singlet bonds between arbitrary spin-$S$ representations of $\qsu$.

\section{\label{sc:HigherSpins}Entanglement of singlet bonds between higher spins}

  In the previous sections we focused on the \qAKLT state for the spin-$1$ representation of $\qsu$. This state is obtained from a valence-bond construction involving two spin-$\frac{1}{2}$ auxiliary spins. It is straightforward to generalize this construction to spin-$S$ auxiliary spins, resulting in a spin-$2S$ analogue of the \qAKLT state.\cite{Motegi:2010PhLA..374.3112M} Just as for the ordinary \qAKLT state, the correlation functions and entanglement properties of these states have been discussed in great detail.\cite{Arita:2011JMP....52f3303A,Santos:2012JPhA...45q5303S,Arita:2012SIGMA...8..081A} As these entanglement considerations were based on periodic boundary conditions which are not compatible with invariance under $\qsu$ we now revisit this issue from an iMPS perspective.

  To keep notation simple, let us consider the singlet bond between two spin-$S$ representations which serves as a higher spin model for the Schmidt decomposition \eqref{eq:SchmidtDecomposition}. Using the explicit action~\eqref{eq:Generators} of the quantum group generators it may easily be verified that the normalized singlet state can be written as
\begin{align}
|\text{singlet}\rangle
=\frac{1}{\sqrt{\qdim(S)}}\sum_{m=-S}^S(-q)^{-m}|m\rangle\otimes|{-}m\rangle\;,
\end{align}
  where the quantum dimension $\qdim(S)$ has been defined in Eq.~\eqref{eq:QuantumDimension}. If $\rho$ denotes the associated density matrix then the $q$-deformed reduced density matrix in this state is given by
\begin{align}
  \rho_R^{(q)}
  =\tr_L\Bigl[q^{2\bS_L^z}\rho\Bigr]
  =\frac{1}{\qdim(S)}\,\idop\;.
\end{align}
  We see that the resulting $q$-deformed entanglement spectrum exhibits a full degeneracy of its $\dim(S)$ $q$-deformed entanglement energies
\begin{align}
  \epsilon_m^{(q)}
  =\log\qdim(S)\;,
\end{align}
  where $m=-S,\ldots,S$. It is then also straightforward to determine the associated $q$-deformed entanglement entropy which is given by
\begin{align}
  S_{EE}^{(q)}
  =\tr_R\Bigl[q^{-2\bS_B^z}\rho_R^{(q)}\log\rho_R^{(q)}\Bigr]
  =\log\qdim(S)\;.
\end{align}
  We recognize that the $q$-deformed entanglement entropy precisely captures the quantum dimension of the two auxiliary spins forming the singlet bond.

\section{\label{sc:Classification}Classification}

  Let us finally adopt a slightly more general perspective. It is known that any gapped ground state $|\psi\rangle$ can be well approximated by means of an MPS.\cite{Hastings:2007JSMTE..08...24H} With $\qsu$ symmetry and integer physical spins there are two distinct classes of MPS, just as for the standard $su(2)$ case. This is due to the fact that the representation theory of $\qsu$ at real values $q\neq0$ precisely mimics the representation theory of $su(2)$, including labeling and dimensions of irreducible representations and tensor product decompositions.\cite{Klimyk:MR1492989}
  
  In particular, integer physical spins can only arise from either two integer or two half-integer auxiliary spins. For $su(2)$, all of these representations lift to $SU(2)$ while only integer spin representations lift to $SO(3)=SU(2)/\Integer_2$. In contrast, half-integer spins are only projective representations of $SO(3)$ since they have a non-trivial action of the central subgroup $\Integer_2\subset SU(2)$. This representation of the center $\Integer_2$ can be interpreted as a topological invariant.\cite{Duivenvoorden:2012arXiv1206.2462D} As discussed in more detail in Ref.~\onlinecite{Quella:2020Draft}, similar statements hold true for the quantum group $\qsu$ which allows to define two associated mathematical structures $SU_q(2)$ and $SO_q(3)$ which should be interpreted as distinct exponentiated versions of $\qsu$.
  
  Just as for the undeformed case, the entanglement in MPS with integer auxiliary spins can be removed while this is not the case for half-integer auxiliary spins if we insist on the preservation of $SO_q(3)$ symmetry. We thus expect a $\Integer_2$-classification of $\qsu$-invariant quantum spin chains based on integer spins, the \qAKLT model being a representative of the non-trivial phase.\footnote{There should be no symmetry-protection if half-integer physical spins are involved.}
  
  We recall from Section~\ref{sc:HigherSpins} that singlet bonds between two spin-$S$ representations lead to a $(2S+1)$-dimensional degeneracy in the $q$-deformed entanglement spectrum. For half-integer $S$ this degeneracy is even while it is odd for integer spins. Just as for the ordinary Haldane phase \cite{Pollmann:PhysRevB.81.064439} there is thus a characteristic entanglement signature of the topologically non-trivial phase which is protected by $SO_q(3)$ symmetry.

\section{Conclusions and Outlook}

  We have presented overwhelming evidence that the \qAKLT model (with $q>0$) should be regarded as a representative of a novel type of SPT phase, protected by the $q$-deformed symmetry $SO_q(3)$. Even though the \qAKLT state on an infinite chain shows no degeneracy in the standard mid-cut entanglement spectrum, a two-fold degeneracy is recovered when defining the reduced density matrix with an appropriate quantum trace. This statement remains true even if the entanglement in the chain is very low (i.e.\ for $q\ll1$ or $q\gg1$). In addition, the state exhibits fractionalized spin-$\frac{1}{2}$ boundary spins and non-trivial string order, as already found in earlier studies.\cite{Totsuka:1994JPhA...27.6443T}
  
  Our findings open many directions of further investigations. It has been shown recently that the \qAKLT model can be obtained from the standard AKLT model by means of a procedure called ``Witten's conjugation''.\cite{Wouters:2020arXiv200512825W} It would be interesting to understand whether this mathematical procedure may also be used to infer conclusions about the topological properties of the \qAKLT state.
  
  Also, some of the models of recent physical interest involve quantum group symmetries with $q$ a root of unity. This is for instance the case for the anyonic chains that have been introduced in Ref.~\onlinecite{Gils:2013PhRvB..87w5120G} and for the abstract classification of topological field theories that arise from intertwiner dynamics.\cite{Dittrich:2013arXiv1311.1798D} In both cases, AKLT-like states are known to exist. It would thus be interesting to investigate whether our results and methods carry over to the case $|q|=1$ and, in particular, roots of unity where the representation theory of $\qsu$ becomes considerably more intricate.\cite{Klimyk:MR1492989} Obviously, it would also be natural to revisit higher spin instances of the \qAKLT model.\cite{Motegi:2010PhLA..374.3112M,Arita:2011JMP....52f3303A,Santos:2012JPhA...45q5303S,Arita:2012SIGMA...8..081A}

  However, the most important and far-reaching question is whether there are other kinds of generalized symmetries that are capable of protecting topological order in 1D or even higher-dimensional systems. In a companion paper \cite{Quella:2020Draft} we show that the ideas presented in this article readily generalize to spin chains with arbitrary quantum group symmetry $\qg$ (with real $q\neq0$) where $\g$ is a finite dimensional simple Lie algebra such as $su(N)$, $so(N)$ or $sp(2N)$. These results of Ref.~\onlinecite{Quella:2020Draft} extend the classification of Ref.~\onlinecite{Duivenvoorden:2012arXiv1206.2462D} which has been established for simple Lie groups.
  
  A similar and hence natural class of symmetries deserving further investigation are elliptic quantum groups, i.e.~two-parameter deformations of Lie algebras. On general grounds one would expect arbitrary Hopf-$\ast$ algebras to be good candidates for generalized symmetries, potentially with additional restrictions on their structure.

  In this context one should also ask about the role of the discrete duality symmetries that were discussed in the main text and that involve the transformation $q\to q^{-1}$. Are these capable of protecting topological phases by themselves and, if yes, how does the resulting classification relate to the one obtained with respect to the continuous quantum group? For the undeformed case these questions were answered in Refs.~\onlinecite{Else:2013arXiv1304.0783E,Duivenvoorden:2013arXiv1304.7234D}.
  
  Since even the examples of generalized symmetries that have been discussed here and in Ref.~\onlinecite{Quella:2020Draft} are not amenable to the standard definition of projective representations a new mathematical framework will need to be developed to describe the topological invariants in full generality. It is likely that this framework will employ tools from non-commutative geometry, the natural generalization of geometric objects to an abstract algebraic setting.

  Finally, thinking of implications of our findings beyond one dimension it should be noted that matrix product operators played a significant role in the treatment of two-dimensional topological phases.\cite{Buerschaper:2014AnPhy.351..447B,Bultinck:2017AnPhy.378..183B} Our findings may help to generalize some of these considerations.

\begin{acknowledgments}
  The author would like to thank Eddy Ardonne, Oliver Buerschaper, John Cardy, Bianca Dittrich, Frank G\"ohmann, Wojciech Kaminski, Andreas Kl\"umper, Deniz Kus, Peter Littelmann, Bruno Nachtergaele, Rafael Nepomechie and Roland Van der Veen for useful discussions. This research was conducted by the Australian Research Council Centre of Excellence for Mathematical and Statistical Frontiers (project number CE140100049) and partially funded by the Australian Government. The author would also like to express his utmost gratitude to his wife Luna. Her continuous support, encouragement and forbearance were essential for being able to complete this research during the unprecedented times of home office and Covid-19 restrictions.
\end{acknowledgments}


\begin{thebibliography}{48}%
\makeatletter
\providecommand \@ifxundefined [1]{%
 \@ifx{#1\undefined}
}%
\providecommand \@ifnum [1]{%
 \ifnum #1\expandafter \@firstoftwo
 \else \expandafter \@secondoftwo
 \fi
}%
\providecommand \@ifx [1]{%
 \ifx #1\expandafter \@firstoftwo
 \else \expandafter \@secondoftwo
 \fi
}%
\providecommand \natexlab [1]{#1}%
\providecommand \enquote  [1]{``#1''}%
\providecommand \bibnamefont  [1]{#1}%
\providecommand \bibfnamefont [1]{#1}%
\providecommand \citenamefont [1]{#1}%
\providecommand \href@noop [0]{\@secondoftwo}%
\providecommand \href [0]{\begingroup \@sanitize@url \@href}%
\providecommand \@href[1]{\@@startlink{#1}\@@href}%
\providecommand \@@href[1]{\endgroup#1\@@endlink}%
\providecommand \@sanitize@url [0]{\catcode `\\12\catcode `\$12\catcode
  `\&12\catcode `\#12\catcode `\^12\catcode `\_12\catcode `\%12\relax}%
\providecommand \@@startlink[1]{}%
\providecommand \@@endlink[0]{}%
\providecommand \url  [0]{\begingroup\@sanitize@url \@url }%
\providecommand \@url [1]{\endgroup\@href {#1}{\urlprefix }}%
\providecommand \urlprefix  [0]{URL }%
\providecommand \Eprint [0]{\href }%
\providecommand \doibase [0]{http://dx.doi.org/}%
\providecommand \selectlanguage [0]{\@gobble}%
\providecommand \bibinfo  [0]{\@secondoftwo}%
\providecommand \bibfield  [0]{\@secondoftwo}%
\providecommand \translation [1]{[#1]}%
\providecommand \BibitemOpen [0]{}%
\providecommand \bibitemStop [0]{}%
\providecommand \bibitemNoStop [0]{.\EOS\space}%
\providecommand \EOS [0]{\spacefactor3000\relax}%
\providecommand \BibitemShut  [1]{\csname bibitem#1\endcsname}%
\let\auto@bib@innerbib\@empty
\bibitem [{\citenamefont {{Pollmann}}\ \emph {et~al.}(2012)\citenamefont
  {{Pollmann}}, \citenamefont {{Berg}}, \citenamefont {{Turner}},\ and\
  \citenamefont {{Oshikawa}}}]{Pollmann:2012PhRvB..85g5125P}%
  \BibitemOpen
  \bibfield  {author} {\bibinfo {author} {\bibfnamefont {F.}~\bibnamefont
  {{Pollmann}}}, \bibinfo {author} {\bibfnamefont {E.}~\bibnamefont {{Berg}}},
  \bibinfo {author} {\bibfnamefont {A.~M.}\ \bibnamefont {{Turner}}}, \ and\
  \bibinfo {author} {\bibfnamefont {M.}~\bibnamefont {{Oshikawa}}},\ }\href
  {\doibase 10.1103/PhysRevB.85.075125} {\bibfield  {journal} {\bibinfo
  {journal} {Phys. Rev.}\ }\textbf {\bibinfo {volume} {B85}},\ \bibinfo {eid}
  {075125} (\bibinfo {year} {2012})},\ \Eprint {http://arxiv.org/abs/0909.4059}
  {arXiv:0909.4059 [cond-mat.str-el]} \BibitemShut {NoStop}%
\bibitem [{\citenamefont {Schuch}\ \emph {et~al.}(2011)\citenamefont {Schuch},
  \citenamefont {P{\'e}rez-Garc{\'i}a},\ and\ \citenamefont
  {Cirac}}]{Schuch:1010.3732v3}%
  \BibitemOpen
  \bibfield  {author} {\bibinfo {author} {\bibfnamefont {N.}~\bibnamefont
  {Schuch}}, \bibinfo {author} {\bibfnamefont {D.}~\bibnamefont
  {P{\'e}rez-Garc{\'i}a}}, \ and\ \bibinfo {author} {\bibfnamefont
  {I.}~\bibnamefont {Cirac}},\ }\href {http://arxiv.org/abs/1010.3732}
  {\bibfield  {journal} {\bibinfo  {journal} {Phys. Rev.}\ }\textbf {\bibinfo
  {volume} {B84}},\ \bibinfo {pages} {165139} (\bibinfo {year} {2011})},\
  \Eprint {http://arxiv.org/abs/1010.3732} {arXiv:1010.3732 [cond-mat.str-el]}
  \BibitemShut {NoStop}%
\bibitem [{\citenamefont {Chen}\ \emph {et~al.}(2011)\citenamefont {Chen},
  \citenamefont {Gu},\ and\ \citenamefont {Wen}}]{Chen:PhysRevB.84.235128}%
  \BibitemOpen
  \bibfield  {author} {\bibinfo {author} {\bibfnamefont {X.}~\bibnamefont
  {Chen}}, \bibinfo {author} {\bibfnamefont {Z.-C.}\ \bibnamefont {Gu}}, \ and\
  \bibinfo {author} {\bibfnamefont {X.-G.}\ \bibnamefont {Wen}},\ }\href
  {\doibase 10.1103/PhysRevB.84.235128} {\bibfield  {journal} {\bibinfo
  {journal} {Phys. Rev.}\ }\textbf {\bibinfo {volume} {B84}},\ \bibinfo {pages}
  {235128} (\bibinfo {year} {2011})},\ \Eprint {http://arxiv.org/abs/1103.3323}
  {arXiv:1103.3323 [cond-mat.str-el]} \BibitemShut {NoStop}%
\bibitem [{\citenamefont {Affleck}\ \emph {et~al.}(1987)\citenamefont
  {Affleck}, \citenamefont {Kennedy}, \citenamefont {Lieb},\ and\ \citenamefont
  {Tasaki}}]{Affleck:PhysRevLett.59.799}%
  \BibitemOpen
  \bibfield  {author} {\bibinfo {author} {\bibfnamefont {I.}~\bibnamefont
  {Affleck}}, \bibinfo {author} {\bibfnamefont {T.}~\bibnamefont {Kennedy}},
  \bibinfo {author} {\bibfnamefont {E.~H.}\ \bibnamefont {Lieb}}, \ and\
  \bibinfo {author} {\bibfnamefont {H.}~\bibnamefont {Tasaki}},\ }\href
  {\doibase 10.1103/PhysRevLett.59.799} {\bibfield  {journal} {\bibinfo
  {journal} {Phys. Rev. Lett.}\ }\textbf {\bibinfo {volume} {59}},\ \bibinfo
  {pages} {799} (\bibinfo {year} {1987})}\BibitemShut {NoStop}%
\bibitem [{\citenamefont {Affleck}\ \emph {et~al.}(1988)\citenamefont
  {Affleck}, \citenamefont {Kennedy}, \citenamefont {Lieb},\ and\ \citenamefont
  {Tasaki}}]{Affleck:1987cy}%
  \BibitemOpen
  \bibfield  {author} {\bibinfo {author} {\bibfnamefont {I.}~\bibnamefont
  {Affleck}}, \bibinfo {author} {\bibfnamefont {T.}~\bibnamefont {Kennedy}},
  \bibinfo {author} {\bibfnamefont {E.~H.}\ \bibnamefont {Lieb}}, \ and\
  \bibinfo {author} {\bibfnamefont {H.}~\bibnamefont {Tasaki}},\ }\href
  {\doibase 10.1007/BF01218021} {\bibfield  {journal} {\bibinfo  {journal}
  {Commun. Math. Phys.}\ }\textbf {\bibinfo {volume} {115}},\ \bibinfo {pages}
  {477} (\bibinfo {year} {1988})}\BibitemShut {NoStop}%
\bibitem [{\citenamefont {den Nijs}\ and\ \citenamefont
  {Rommelse}(1989)}]{DenNijs:PhysRevB.40.4709}%
  \BibitemOpen
  \bibfield  {author} {\bibinfo {author} {\bibfnamefont {M.}~\bibnamefont {den
  Nijs}}\ and\ \bibinfo {author} {\bibfnamefont {K.}~\bibnamefont {Rommelse}},\
  }\href {\doibase 10.1103/PhysRevB.40.4709} {\bibfield  {journal} {\bibinfo
  {journal} {Phys. Rev.}\ }\textbf {\bibinfo {volume} {B40}},\ \bibinfo {pages}
  {4709} (\bibinfo {year} {1989})}\BibitemShut {NoStop}%
\bibitem [{\citenamefont {Kennedy}\ and\ \citenamefont
  {Tasaki}(1992)}]{Kennedy:10.1007/BF02097239}%
  \BibitemOpen
  \bibfield  {author} {\bibinfo {author} {\bibfnamefont {T.}~\bibnamefont
  {Kennedy}}\ and\ \bibinfo {author} {\bibfnamefont {H.}~\bibnamefont
  {Tasaki}},\ }\href {http://dx.doi.org/10.1007/BF02097239} {\bibfield
  {journal} {\bibinfo  {journal} {Commun. Math. Phys.}\ }\textbf {\bibinfo
  {volume} {147}},\ \bibinfo {pages} {431} (\bibinfo {year}
  {1992})}\BibitemShut {NoStop}%
\bibitem [{\citenamefont {Pollmann}\ \emph {et~al.}(2010)\citenamefont
  {Pollmann}, \citenamefont {Turner}, \citenamefont {Berg},\ and\ \citenamefont
  {Oshikawa}}]{Pollmann:PhysRevB.81.064439}%
  \BibitemOpen
  \bibfield  {author} {\bibinfo {author} {\bibfnamefont {F.}~\bibnamefont
  {Pollmann}}, \bibinfo {author} {\bibfnamefont {A.~M.}\ \bibnamefont
  {Turner}}, \bibinfo {author} {\bibfnamefont {E.}~\bibnamefont {Berg}}, \ and\
  \bibinfo {author} {\bibfnamefont {M.}~\bibnamefont {Oshikawa}},\ }\href
  {\doibase 10.1103/PhysRevB.81.064439} {\bibfield  {journal} {\bibinfo
  {journal} {Phys. Rev.}\ }\textbf {\bibinfo {volume} {B81}},\ \bibinfo {pages}
  {064439} (\bibinfo {year} {2010})},\ \Eprint {http://arxiv.org/abs/0910.1811}
  {arXiv:0910.1811 [cond-mat.str-el]} \BibitemShut {NoStop}%
\bibitem [{\citenamefont {Fannes}\ \emph {et~al.}(1989)\citenamefont {Fannes},
  \citenamefont {Nachtergaele},\ and\ \citenamefont {Werner}}]{Fannes:1989ns}%
  \BibitemOpen
  \bibfield  {author} {\bibinfo {author} {\bibfnamefont {M.}~\bibnamefont
  {Fannes}}, \bibinfo {author} {\bibfnamefont {B.}~\bibnamefont
  {Nachtergaele}}, \ and\ \bibinfo {author} {\bibfnamefont {R.~F.}\
  \bibnamefont {Werner}},\ }\href {\doibase 10.1209/0295-5075/10/7/005}
  {\bibfield  {journal} {\bibinfo  {journal} {Europhys. Lett.}\ }\textbf
  {\bibinfo {volume} {10}},\ \bibinfo {pages} {633} (\bibinfo {year}
  {1989})}\BibitemShut {NoStop}%
\bibitem [{\citenamefont {Fannes}\ \emph {et~al.}(1991)\citenamefont {Fannes},
  \citenamefont {Nachtergaele},\ and\ \citenamefont {Werner}}]{Fannes:1990px}%
  \BibitemOpen
  \bibfield  {author} {\bibinfo {author} {\bibfnamefont {M.~B.}\ \bibnamefont
  {Fannes}}, \bibinfo {author} {\bibfnamefont {B.}~\bibnamefont
  {Nachtergaele}}, \ and\ \bibinfo {author} {\bibfnamefont {R.~F.}\
  \bibnamefont {Werner}},\ }\href@noop {} {\bibfield  {journal} {\bibinfo
  {journal} {J. Phys.}\ }\textbf {\bibinfo {volume} {A24}},\ \bibinfo {pages}
  {L185} (\bibinfo {year} {1991})}\BibitemShut {NoStop}%
\bibitem [{\citenamefont {Fannes}\ \emph {et~al.}(1992)\citenamefont {Fannes},
  \citenamefont {Nachtergaele},\ and\ \citenamefont {Werner}}]{Fannes:1990ur}%
  \BibitemOpen
  \bibfield  {author} {\bibinfo {author} {\bibfnamefont {M.}~\bibnamefont
  {Fannes}}, \bibinfo {author} {\bibfnamefont {B.}~\bibnamefont
  {Nachtergaele}}, \ and\ \bibinfo {author} {\bibfnamefont {R.~F.}\
  \bibnamefont {Werner}},\ }\href {\doibase 10.1007/BF02099178} {\bibfield
  {journal} {\bibinfo  {journal} {Commun. Math. Phys.}\ }\textbf {\bibinfo
  {volume} {144}},\ \bibinfo {pages} {443} (\bibinfo {year}
  {1992})}\BibitemShut {NoStop}%
\bibitem [{\citenamefont {P{\'e}rez-Garc{\'i}a}\ \emph
  {et~al.}(2007)\citenamefont {P{\'e}rez-Garc{\'i}a}, \citenamefont
  {Verstraete}, \citenamefont {Wolf},\ and\ \citenamefont
  {Cirac}}]{Perez-Garcia:2007:MPS:2011832.2011833}%
  \BibitemOpen
  \bibfield  {author} {\bibinfo {author} {\bibfnamefont {D.}~\bibnamefont
  {P{\'e}rez-Garc{\'i}a}}, \bibinfo {author} {\bibfnamefont {F.}~\bibnamefont
  {Verstraete}}, \bibinfo {author} {\bibfnamefont {M.~M.}\ \bibnamefont
  {Wolf}}, \ and\ \bibinfo {author} {\bibfnamefont {J.~I.}\ \bibnamefont
  {Cirac}},\ }\href {http://dl.acm.org/citation.cfm?id=2011832.2011833}
  {\bibfield  {journal} {\bibinfo  {journal} {Quantum Info. Comput.}\ }\textbf
  {\bibinfo {volume} {7}},\ \bibinfo {pages} {401} (\bibinfo {year} {2007})},\
  \Eprint {http://arxiv.org/abs/arXiv:quant-ph/0608197}
  {arXiv:quant-ph/0608197} \BibitemShut {NoStop}%
\bibitem [{\citenamefont
  {{Schollw{\"o}ck}}(2011)}]{Schollwock:2011AnPhy.326...96S}%
  \BibitemOpen
  \bibfield  {author} {\bibinfo {author} {\bibfnamefont {U.}~\bibnamefont
  {{Schollw{\"o}ck}}},\ }\href {\doibase 10.1016/j.aop.2010.09.012} {\bibfield
  {journal} {\bibinfo  {journal} {Annals Phys.}\ }\textbf {\bibinfo {volume}
  {326}},\ \bibinfo {pages} {96} (\bibinfo {year} {2011})},\ \Eprint
  {http://arxiv.org/abs/1008.3477} {arXiv:1008.3477 [cond-mat.str-el]}
  \BibitemShut {NoStop}%
\bibitem [{\citenamefont {{Batchelor}}\ \emph {et~al.}(1990)\citenamefont
  {{Batchelor}}, \citenamefont {{Mezincescu}}, \citenamefont {{Nepomechie}},\
  and\ \citenamefont {{Rittenberg}}}]{Batchelor:1990JPhA...23L.141B}%
  \BibitemOpen
  \bibfield  {author} {\bibinfo {author} {\bibfnamefont {M.~T.}\ \bibnamefont
  {{Batchelor}}}, \bibinfo {author} {\bibfnamefont {L.}~\bibnamefont
  {{Mezincescu}}}, \bibinfo {author} {\bibfnamefont {R.~I.}\ \bibnamefont
  {{Nepomechie}}}, \ and\ \bibinfo {author} {\bibfnamefont {V.}~\bibnamefont
  {{Rittenberg}}},\ }\href {\doibase 10.1088/0305-4470/23/4/003} {\bibfield
  {journal} {\bibinfo  {journal} {J. Phys.}\ }\textbf {\bibinfo {volume}
  {A23}},\ \bibinfo {pages} {L141} (\bibinfo {year} {1990})}\BibitemShut
  {NoStop}%
\bibitem [{\citenamefont {{Kl{\"u}mper}}\ \emph {et~al.}(1991)\citenamefont
  {{Kl{\"u}mper}}, \citenamefont {{Schadschneider}},\ and\ \citenamefont
  {{Zittartz}}}]{Klumper:1991JPhA...24L.955K}%
  \BibitemOpen
  \bibfield  {author} {\bibinfo {author} {\bibfnamefont {A.}~\bibnamefont
  {{Kl{\"u}mper}}}, \bibinfo {author} {\bibfnamefont {A.}~\bibnamefont
  {{Schadschneider}}}, \ and\ \bibinfo {author} {\bibfnamefont
  {J.}~\bibnamefont {{Zittartz}}},\ }\href {\doibase
  10.1088/0305-4470/24/16/012} {\bibfield  {journal} {\bibinfo  {journal} {J.
  Phys.}\ }\textbf {\bibinfo {volume} {A24}},\ \bibinfo {pages} {L955}
  (\bibinfo {year} {1991})}\BibitemShut {NoStop}%
\bibitem [{\citenamefont {{Kl{\"u}mper}}\ \emph {et~al.}(1992)\citenamefont
  {{Kl{\"u}mper}}, \citenamefont {{Schadschneider}},\ and\ \citenamefont
  {{Zittartz}}}]{Klumper:1992ZPhyB..87..281K}%
  \BibitemOpen
  \bibfield  {author} {\bibinfo {author} {\bibfnamefont {A.}~\bibnamefont
  {{Kl{\"u}mper}}}, \bibinfo {author} {\bibfnamefont {A.}~\bibnamefont
  {{Schadschneider}}}, \ and\ \bibinfo {author} {\bibfnamefont
  {J.}~\bibnamefont {{Zittartz}}},\ }\href {\doibase 10.1007/BF01309281}
  {\bibfield  {journal} {\bibinfo  {journal} {Zeit. Phys.}\ }\textbf {\bibinfo
  {volume} {B87}},\ \bibinfo {pages} {281} (\bibinfo {year}
  {1992})}\BibitemShut {NoStop}%
\bibitem [{\citenamefont {Batchelor}\ and\ \citenamefont
  {Yung}(1994)}]{Batchelor:1994IJMPB...8.3645B}%
  \BibitemOpen
  \bibfield  {author} {\bibinfo {author} {\bibfnamefont {M.~T.}\ \bibnamefont
  {Batchelor}}\ and\ \bibinfo {author} {\bibfnamefont {C.~M.}\ \bibnamefont
  {Yung}},\ }\href {\doibase 10.1142/S021797929400155X} {\bibfield  {journal}
  {\bibinfo  {journal} {Int. J. Mod. Phys.}\ }\textbf {\bibinfo {volume}
  {B8}},\ \bibinfo {pages} {3645} (\bibinfo {year} {1994})},\ \Eprint
  {http://arxiv.org/abs/arXiv:cond-mat/9403080} {arXiv:cond-mat/9403080}
  \BibitemShut {NoStop}%
\bibitem [{\citenamefont {{Totsuka}}\ and\ \citenamefont
  {{Suzuki}}(1994)}]{Totsuka:1994JPhA...27.6443T}%
  \BibitemOpen
  \bibfield  {author} {\bibinfo {author} {\bibfnamefont {K.}~\bibnamefont
  {{Totsuka}}}\ and\ \bibinfo {author} {\bibfnamefont {M.}~\bibnamefont
  {{Suzuki}}},\ }\href {\doibase 10.1088/0305-4470/27/19/017} {\bibfield
  {journal} {\bibinfo  {journal} {J. Phys.}\ }\textbf {\bibinfo {volume}
  {A27}},\ \bibinfo {pages} {6443} (\bibinfo {year} {1994})}\BibitemShut
  {NoStop}%
\bibitem [{\citenamefont {Fannes}\ \emph {et~al.}(1996)\citenamefont {Fannes},
  \citenamefont {Nachtergaele},\ and\ \citenamefont
  {Werner}}]{Fannes:10.1007/BF02101525}%
  \BibitemOpen
  \bibfield  {author} {\bibinfo {author} {\bibfnamefont {M.}~\bibnamefont
  {Fannes}}, \bibinfo {author} {\bibfnamefont {B.}~\bibnamefont
  {Nachtergaele}}, \ and\ \bibinfo {author} {\bibfnamefont {R.}~\bibnamefont
  {Werner}},\ }\href {http://dx.doi.org/10.1007/BF02101525} {\bibfield
  {journal} {\bibinfo  {journal} {Commun. Math. Phys.}\ }\textbf {\bibinfo
  {volume} {174}},\ \bibinfo {pages} {477} (\bibinfo {year} {1996})},\ \Eprint
  {http://arxiv.org/abs/arXiv:cond-mat/9504002} {arXiv:cond-mat/9504002}
  \BibitemShut {NoStop}%
\bibitem [{\citenamefont {{Santos}}\ \emph
  {et~al.}(2012{\natexlab{a}})\citenamefont {{Santos}}, \citenamefont
  {{Paraan}}, \citenamefont {{Korepin}},\ and\ \citenamefont
  {{Kl{\"u}mper}}}]{Santos:2012EL.....9837005S}%
  \BibitemOpen
  \bibfield  {author} {\bibinfo {author} {\bibfnamefont {R.~A.}\ \bibnamefont
  {{Santos}}}, \bibinfo {author} {\bibfnamefont {F.~N.~C.}\ \bibnamefont
  {{Paraan}}}, \bibinfo {author} {\bibfnamefont {V.~E.}\ \bibnamefont
  {{Korepin}}}, \ and\ \bibinfo {author} {\bibfnamefont {A.}~\bibnamefont
  {{Kl{\"u}mper}}},\ }\href {\doibase 10.1209/0295-5075/98/37005} {\bibfield
  {journal} {\bibinfo  {journal} {EPL (Europhysics Letters)}\ }\textbf
  {\bibinfo {volume} {98}},\ \bibinfo {pages} {37005} (\bibinfo {year}
  {2012}{\natexlab{a}})},\ \Eprint {http://arxiv.org/abs/1112.0517}
  {arXiv:1112.0517 [quant-ph]} \BibitemShut {NoStop}%
\bibitem [{Note1()}]{Note1}%
  \BibitemOpen
  \bibinfo {note} {A connection between quantum group symmetries and SPT phases
  was already anticipated in Ref.~\protect \rev@citealpnum
  {Duivenvoorden:2012arXiv1206.2462D}. Speculations about the $q\protect \text
  {AKLT}$ model (and its generalizations to higher spin) realizing an SPT phase
  have also appeared in Ref.~\protect \rev@citealpnum
  {Dittrich:2013arXiv1311.1798D}. However, the latter suggestion seems to boil
  down to a pure analogy.}\BibitemShut {Stop}%
\bibitem [{Note2()}]{Note2}%
  \BibitemOpen
  \bibinfo {note} {We could say $SU(2)$ here since $SO(3)=SU(2)/\protect
  \mathbb {Z}_2$ and the $\protect \mathbb {Z}_2$ subgroup acts trivially on
  integer spins. However, from a more general perspective it is more
  appropriate to think of $SO(3)$ as the protecting symmetry\cite
  {Duivenvoorden:2012arXiv1206.2462D}}\BibitemShut {NoStop}%
\bibitem [{Note3()}]{Note3}%
  \BibitemOpen
  \bibinfo {note} {Our conventions are based on what is called $\protect \breve
  {\protect \mathcal {U}}_q(sl_2)$ in Ref.~\protect \rev@citealpnum
  {Klimyk:MR1492989}. We note that Ref.~\protect \rev@citealpnum
  {Klimyk:MR1492989} has reserved the symbol $\protect \mathcal {U}_q(sl_2)$ to
  denote the same quantum group with a different coproduct. It is worth
  pointing out that only the choice of coproduct used in this article is
  consistent with the usual hermiticity properties of spins.}\BibitemShut
  {Stop}%
\bibitem [{\citenamefont {{Pasquier}}\ and\ \citenamefont
  {{Saleur}}(1990)}]{Pasquier:1990NuPhB.330..523P}%
  \BibitemOpen
  \bibfield  {author} {\bibinfo {author} {\bibfnamefont {V.}~\bibnamefont
  {{Pasquier}}}\ and\ \bibinfo {author} {\bibfnamefont {H.}~\bibnamefont
  {{Saleur}}},\ }\href {\doibase 10.1016/0550-3213(90)90122-T} {\bibfield
  {journal} {\bibinfo  {journal} {Nucl. Phys.}\ }\textbf {\bibinfo {volume}
  {B330}},\ \bibinfo {pages} {523} (\bibinfo {year} {1990})}\BibitemShut
  {NoStop}%
\bibitem [{\citenamefont {{Karowski}}\ and\ \citenamefont
  {{Zapletal}}(1994)}]{Karowski:1994NuPhB.419..567K}%
  \BibitemOpen
  \bibfield  {author} {\bibinfo {author} {\bibfnamefont {M.}~\bibnamefont
  {{Karowski}}}\ and\ \bibinfo {author} {\bibfnamefont {A.}~\bibnamefont
  {{Zapletal}}},\ }\href {\doibase 10.1016/0550-3213(94)90345-X} {\bibfield
  {journal} {\bibinfo  {journal} {Nucl. Phys.}\ }\textbf {\bibinfo {volume}
  {B419}},\ \bibinfo {pages} {567} (\bibinfo {year} {1994})},\ \Eprint
  {http://arxiv.org/abs/hep-th/9312008} {arXiv:hep-th/9312008 [hep-th]}
  \BibitemShut {NoStop}%
\bibitem [{\citenamefont {{Grosse}}\ \emph {et~al.}(1994)\citenamefont
  {{Grosse}}, \citenamefont {{Pallua}}, \citenamefont {{Prester}},\ and\
  \citenamefont {{Raschhofer}}}]{Grosse:1994JPhA...27.4761G}%
  \BibitemOpen
  \bibfield  {author} {\bibinfo {author} {\bibfnamefont {H.}~\bibnamefont
  {{Grosse}}}, \bibinfo {author} {\bibfnamefont {S.}~\bibnamefont {{Pallua}}},
  \bibinfo {author} {\bibfnamefont {P.}~\bibnamefont {{Prester}}}, \ and\
  \bibinfo {author} {\bibfnamefont {E.}~\bibnamefont {{Raschhofer}}},\ }\href
  {\doibase 10.1088/0305-4470/27/14/007} {\bibfield  {journal} {\bibinfo
  {journal} {Journal of Physics A Mathematical General}\ }\textbf {\bibinfo
  {volume} {27}},\ \bibinfo {pages} {4761} (\bibinfo {year}
  {1994})}\BibitemShut {NoStop}%
\bibitem [{\citenamefont {{Links}}\ \emph {et~al.}(1999)\citenamefont
  {{Links}}, \citenamefont {{Foerster}},\ and\ \citenamefont
  {{Karowski}}}]{Links:1999JMP....40..726L}%
  \BibitemOpen
  \bibfield  {author} {\bibinfo {author} {\bibfnamefont {J.}~\bibnamefont
  {{Links}}}, \bibinfo {author} {\bibfnamefont {A.}~\bibnamefont {{Foerster}}},
  \ and\ \bibinfo {author} {\bibfnamefont {M.}~\bibnamefont {{Karowski}}},\
  }\href {\doibase 10.1063/1.532701} {\bibfield  {journal} {\bibinfo  {journal}
  {Journal of Mathematical Physics}\ }\textbf {\bibinfo {volume} {40}},\
  \bibinfo {pages} {726} (\bibinfo {year} {1999})},\ \Eprint
  {http://arxiv.org/abs/solv-int/9809001} {arXiv:solv-int/9809001 [nlin.SI]}
  \BibitemShut {NoStop}%
\bibitem [{Note4()}]{Note4}%
  \BibitemOpen
  \bibinfo {note} {We note that the combination of two of these tansformations
  is still a symmetry of the Hamiltonian~\protect \textup {\hbox {\mathsurround
  \z@ \protect \normalfont (\ignorespaces \ref {eq:qAKLTHamiltonian}\unskip
  \@@italiccorr )}}. However, inversion symmetry for instance can easily be
  broken by staggering the couplings without changing the ground state or its
  topological properties.}\BibitemShut {Stop}%
\bibitem [{\citenamefont {{Vidal}}(2007)}]{Vidal:2007PhRvL..98g0201V}%
  \BibitemOpen
  \bibfield  {author} {\bibinfo {author} {\bibfnamefont {G.}~\bibnamefont
  {{Vidal}}},\ }\href {\doibase 10.1103/PhysRevLett.98.070201} {\bibfield
  {journal} {\bibinfo  {journal} {Phys. Rev. Lett.}\ }\textbf {\bibinfo
  {volume} {98}},\ \bibinfo {eid} {070201} (\bibinfo {year} {2007})},\ \Eprint
  {http://arxiv.org/abs/cond-mat/0605597} {arXiv:cond-mat/0605597
  [cond-mat.str-el]} \BibitemShut {NoStop}%
\bibitem [{\citenamefont {Or\'us}\ and\ \citenamefont
  {Vidal}(2008)}]{Orus:PhysRevB.78.155117}%
  \BibitemOpen
  \bibfield  {author} {\bibinfo {author} {\bibfnamefont {R.}~\bibnamefont
  {Or\'us}}\ and\ \bibinfo {author} {\bibfnamefont {G.}~\bibnamefont {Vidal}},\
  }\href {\doibase 10.1103/PhysRevB.78.155117} {\bibfield  {journal} {\bibinfo
  {journal} {Phys. Rev.}\ }\textbf {\bibinfo {volume} {B78}},\ \bibinfo {pages}
  {155117} (\bibinfo {year} {2008})}\BibitemShut {NoStop}%
\bibitem [{\citenamefont {{Couvreur}}\ \emph {et~al.}(2017)\citenamefont
  {{Couvreur}}, \citenamefont {{Jacobsen}},\ and\ \citenamefont
  {{Saleur}}}]{Couvreur:2017PhRvL.119d0601C}%
  \BibitemOpen
  \bibfield  {author} {\bibinfo {author} {\bibfnamefont {R.}~\bibnamefont
  {{Couvreur}}}, \bibinfo {author} {\bibfnamefont {J.~L.}\ \bibnamefont
  {{Jacobsen}}}, \ and\ \bibinfo {author} {\bibfnamefont {H.}~\bibnamefont
  {{Saleur}}},\ }\href {\doibase 10.1103/PhysRevLett.119.040601} {\bibfield
  {journal} {\bibinfo  {journal} {Phys. Rev. Lett.}\ }\textbf {\bibinfo
  {volume} {119}},\ \bibinfo {eid} {040601} (\bibinfo {year} {2017})},\ \Eprint
  {http://arxiv.org/abs/1611.08506} {arXiv:1611.08506 [cond-mat.stat-mech]}
  \BibitemShut {NoStop}%
\bibitem [{\citenamefont {{Motegi}}(2010)}]{Motegi:2010PhLA..374.3112M}%
  \BibitemOpen
  \bibfield  {author} {\bibinfo {author} {\bibfnamefont {K.}~\bibnamefont
  {{Motegi}}},\ }\href {\doibase 10.1016/j.physleta.2010.05.055} {\bibfield
  {journal} {\bibinfo  {journal} {Phys. Lett.}\ }\textbf {\bibinfo {volume}
  {A374}},\ \bibinfo {pages} {3112} (\bibinfo {year} {2010})},\ \Eprint
  {http://arxiv.org/abs/1003.0050} {arXiv:1003.0050 [math-ph]} \BibitemShut
  {NoStop}%
\bibitem [{\citenamefont {{Arita}}\ and\ \citenamefont
  {{Motegi}}(2011)}]{Arita:2011JMP....52f3303A}%
  \BibitemOpen
  \bibfield  {author} {\bibinfo {author} {\bibfnamefont {C.}~\bibnamefont
  {{Arita}}}\ and\ \bibinfo {author} {\bibfnamefont {K.}~\bibnamefont
  {{Motegi}}},\ }\href {\doibase 10.1063/1.3598424} {\bibfield  {journal}
  {\bibinfo  {journal} {J. Math. Phys.}\ }\textbf {\bibinfo {volume} {52}},\
  \bibinfo {pages} {063303} (\bibinfo {year} {2011})},\ \Eprint
  {http://arxiv.org/abs/1009.4018} {arXiv:1009.4018 [math-ph]} \BibitemShut
  {NoStop}%
\bibitem [{\citenamefont {{Santos}}\ \emph
  {et~al.}(2012{\natexlab{b}})\citenamefont {{Santos}}, \citenamefont
  {{Paraan}}, \citenamefont {{Korepin}},\ and\ \citenamefont
  {{Kl{\"u}mper}}}]{Santos:2012JPhA...45q5303S}%
  \BibitemOpen
  \bibfield  {author} {\bibinfo {author} {\bibfnamefont {R.~A.}\ \bibnamefont
  {{Santos}}}, \bibinfo {author} {\bibfnamefont {F.~N.~C.}\ \bibnamefont
  {{Paraan}}}, \bibinfo {author} {\bibfnamefont {V.~E.}\ \bibnamefont
  {{Korepin}}}, \ and\ \bibinfo {author} {\bibfnamefont {A.}~\bibnamefont
  {{Kl{\"u}mper}}},\ }\href {\doibase 10.1088/1751-8113/45/17/175303}
  {\bibfield  {journal} {\bibinfo  {journal} {J. Phys.}\ }\textbf {\bibinfo
  {volume} {A45}},\ \bibinfo {eid} {175303} (\bibinfo {year}
  {2012}{\natexlab{b}})},\ \Eprint {http://arxiv.org/abs/1201.5927}
  {arXiv:1201.5927 [quant-ph]} \BibitemShut {NoStop}%
\bibitem [{\citenamefont {{Arita}}\ and\ \citenamefont
  {{Motegi}}(2012)}]{Arita:2012SIGMA...8..081A}%
  \BibitemOpen
  \bibfield  {author} {\bibinfo {author} {\bibfnamefont {C.}~\bibnamefont
  {{Arita}}}\ and\ \bibinfo {author} {\bibfnamefont {K.}~\bibnamefont
  {{Motegi}}},\ }\href {\doibase 10.3842/SIGMA.2012.081} {\bibfield  {journal}
  {\bibinfo  {journal} {SIGMA}\ }\textbf {\bibinfo {volume} {8}},\ \bibinfo
  {eid} {081} (\bibinfo {year} {2012})},\ \Eprint
  {http://arxiv.org/abs/1206.3653} {arXiv:1206.3653 [cond-mat.stat-mech]}
  \BibitemShut {NoStop}%
\bibitem [{\citenamefont {{Hastings}}(2007)}]{Hastings:2007JSMTE..08...24H}%
  \BibitemOpen
  \bibfield  {author} {\bibinfo {author} {\bibfnamefont {M.~B.}\ \bibnamefont
  {{Hastings}}},\ }\href {\doibase 10.1088/1742-5468/2007/08/P08024} {\bibfield
   {journal} {\bibinfo  {journal} {J. Stat. Mech.}\ }\textbf {\bibinfo {volume}
  {8}},\ \bibinfo {pages} {24} (\bibinfo {year} {2007})},\ \Eprint
  {http://arxiv.org/abs/0705.2024} {arXiv:0705.2024 [quant-ph]} \BibitemShut
  {NoStop}%
\bibitem [{\citenamefont {Klimyk}\ and\ \citenamefont
  {Schm\"{u}dgen}(1997)}]{Klimyk:MR1492989}%
  \BibitemOpen
  \bibfield  {author} {\bibinfo {author} {\bibfnamefont {A.}~\bibnamefont
  {Klimyk}}\ and\ \bibinfo {author} {\bibfnamefont {K.}~\bibnamefont
  {Schm\"{u}dgen}},\ }\href {\doibase 10.1007/978-3-642-60896-4} {\emph
  {\bibinfo {title} {Quantum groups and their representations}}},\ Texts and
  Monographs in Physics\ (\bibinfo  {publisher} {Springer-Verlag, Berlin},\
  \bibinfo {year} {1997})\ pp.\ \bibinfo {pages} {xx+552}\BibitemShut {NoStop}%
\bibitem [{\citenamefont {{Duivenvoorden}}\ and\ \citenamefont
  {{Quella}}(2013{\natexlab{a}})}]{Duivenvoorden:2012arXiv1206.2462D}%
  \BibitemOpen
  \bibfield  {author} {\bibinfo {author} {\bibfnamefont {K.}~\bibnamefont
  {{Duivenvoorden}}}\ and\ \bibinfo {author} {\bibfnamefont {T.}~\bibnamefont
  {{Quella}}},\ }\href {\doibase 10.1103/PhysRevB.87.125145} {\bibfield
  {journal} {\bibinfo  {journal} {Phys. Rev.}\ }\textbf {\bibinfo {volume}
  {B87}},\ \bibinfo {pages} {125145} (\bibinfo {year} {2013}{\natexlab{a}})},\
  \Eprint {http://arxiv.org/abs/1206.2462} {arXiv:1206.2462 [cond-mat.str-el]}
  \BibitemShut {NoStop}%
\bibitem [{\citenamefont {Quella}()}]{Quella:2020Draft}%
  \BibitemOpen
  \bibfield  {author} {\bibinfo {author} {\bibfnamefont {T.}~\bibnamefont
  {Quella}},\ }\href@noop {} {\enquote {\bibinfo {title} {Symmetry protected
  topological phases beyond groups: $q$-deformed symmetries}, in preparation}\ }\BibitemShut {NoStop}%
\bibitem [{Note5()}]{Note5}%
  \BibitemOpen
  \bibinfo {note} {There should be no symmetry-protection if half-integer
  physical spins are involved.}\BibitemShut {Stop}%
\bibitem [{\citenamefont {{Wouters}}\ \emph {et~al.}(2020)\citenamefont
  {{Wouters}}, \citenamefont {{Katsura}},\ and\ \citenamefont
  {{Schuricht}}}]{Wouters:2020arXiv200512825W}%
  \BibitemOpen
  \bibfield  {author} {\bibinfo {author} {\bibfnamefont {J.}~\bibnamefont
  {{Wouters}}}, \bibinfo {author} {\bibfnamefont {H.}~\bibnamefont
  {{Katsura}}}, \ and\ \bibinfo {author} {\bibfnamefont {D.}~\bibnamefont
  {{Schuricht}}},\ }\href@noop {} {\bibfield  {journal} {\bibinfo  {journal}
  {arXiv e-prints}\ ,\ \bibinfo {eid} {arXiv:2005.12825}} (\bibinfo {year}
  {2020})},\ \Eprint {http://arxiv.org/abs/2005.12825} {arXiv:2005.12825
  [cond-mat.str-el]} \BibitemShut {NoStop}%
\bibitem [{\citenamefont {{Gils}}\ \emph {et~al.}(2013)\citenamefont {{Gils}},
  \citenamefont {{Ardonne}}, \citenamefont {{Trebst}}, \citenamefont {{Huse}},
  \citenamefont {{Ludwig}}, \citenamefont {{Troyer}},\ and\ \citenamefont
  {{Wang}}}]{Gils:2013PhRvB..87w5120G}%
  \BibitemOpen
  \bibfield  {author} {\bibinfo {author} {\bibfnamefont {C.}~\bibnamefont
  {{Gils}}}, \bibinfo {author} {\bibfnamefont {E.}~\bibnamefont {{Ardonne}}},
  \bibinfo {author} {\bibfnamefont {S.}~\bibnamefont {{Trebst}}}, \bibinfo
  {author} {\bibfnamefont {D.~A.}\ \bibnamefont {{Huse}}}, \bibinfo {author}
  {\bibfnamefont {A.~W.~W.}\ \bibnamefont {{Ludwig}}}, \bibinfo {author}
  {\bibfnamefont {M.}~\bibnamefont {{Troyer}}}, \ and\ \bibinfo {author}
  {\bibfnamefont {Z.}~\bibnamefont {{Wang}}},\ }\href {\doibase
  10.1103/PhysRevB.87.235120} {\bibfield  {journal} {\bibinfo  {journal} {Phys.
  Rev.}\ }\textbf {\bibinfo {volume} {B87}},\ \bibinfo {eid} {235120} (\bibinfo
  {year} {2013})},\ \Eprint {http://arxiv.org/abs/1303.4290} {arXiv:1303.4290
  [cond-mat.str-el]} \BibitemShut {NoStop}%
\bibitem [{\citenamefont {{Dittrich}}\ and\ \citenamefont
  {{Kaminski}}(2013)}]{Dittrich:2013arXiv1311.1798D}%
  \BibitemOpen
  \bibfield  {author} {\bibinfo {author} {\bibfnamefont {B.}~\bibnamefont
  {{Dittrich}}}\ and\ \bibinfo {author} {\bibfnamefont {W.}~\bibnamefont
  {{Kaminski}}},\ }\href@noop {} {\bibfield  {journal} {\bibinfo  {journal}
  {arXiv e-prints}\ ,\ \bibinfo {eid} {arXiv:1311.1798}} (\bibinfo {year}
  {2013})},\ \Eprint {http://arxiv.org/abs/1311.1798} {arXiv:1311.1798 [gr-qc]}
  \BibitemShut {NoStop}%
\bibitem [{\citenamefont {{Else}}\ \emph {et~al.}(2013)\citenamefont {{Else}},
  \citenamefont {{Bartlett}},\ and\ \citenamefont
  {{Doherty}}}]{Else:2013arXiv1304.0783E}%
  \BibitemOpen
  \bibfield  {author} {\bibinfo {author} {\bibfnamefont {D.~V.}\ \bibnamefont
  {{Else}}}, \bibinfo {author} {\bibfnamefont {S.~D.}\ \bibnamefont
  {{Bartlett}}}, \ and\ \bibinfo {author} {\bibfnamefont {A.~C.}\ \bibnamefont
  {{Doherty}}},\ }\href {\doibase 10.1103/PhysRevB.88.085114} {\bibfield
  {journal} {\bibinfo  {journal} {Phys. Rev.}\ }\textbf {\bibinfo {volume}
  {B88}},\ \bibinfo {eid} {085114} (\bibinfo {year} {2013})},\ \Eprint
  {http://arxiv.org/abs/1304.0783} {arXiv:1304.0783 [cond-mat.str-el]}
  \BibitemShut {NoStop}%
\bibitem [{\citenamefont {{Duivenvoorden}}\ and\ \citenamefont
  {{Quella}}(2013{\natexlab{b}})}]{Duivenvoorden:2013arXiv1304.7234D}%
  \BibitemOpen
  \bibfield  {author} {\bibinfo {author} {\bibfnamefont {K.}~\bibnamefont
  {{Duivenvoorden}}}\ and\ \bibinfo {author} {\bibfnamefont {T.}~\bibnamefont
  {{Quella}}},\ }\href@noop {} {\bibfield  {journal} {\bibinfo  {journal}
  {Phys. Rev.}\ }\textbf {\bibinfo {volume} {B88}},\ \bibinfo {pages} {125115}
  (\bibinfo {year} {2013}{\natexlab{b}})},\ \Eprint
  {http://arxiv.org/abs/1304.7234} {arXiv:1304.7234 [cond-mat.str-el]}
  \BibitemShut {NoStop}%
\bibitem [{\citenamefont
  {{Buerschaper}}(2014)}]{Buerschaper:2014AnPhy.351..447B}%
  \BibitemOpen
  \bibfield  {author} {\bibinfo {author} {\bibfnamefont {O.}~\bibnamefont
  {{Buerschaper}}},\ }\href {\doibase 10.1016/j.aop.2014.09.007} {\bibfield
  {journal} {\bibinfo  {journal} {Annals Phys.}\ }\textbf {\bibinfo {volume}
  {351}},\ \bibinfo {pages} {447} (\bibinfo {year} {2014})},\ \Eprint
  {http://arxiv.org/abs/1307.7763} {arXiv:1307.7763 [cond-mat.str-el]}
  \BibitemShut {NoStop}%
\bibitem [{\citenamefont {{Bultinck}}\ \emph {et~al.}(2017)\citenamefont
  {{Bultinck}}, \citenamefont {{Mari{\"e}n}}, \citenamefont {{Williamson}},
  \citenamefont {{{\c{S}}ahino{\u{g}}lu}}, \citenamefont {{Haegeman}},\ and\
  \citenamefont {{Verstraete}}}]{Bultinck:2017AnPhy.378..183B}%
  \BibitemOpen
  \bibfield  {author} {\bibinfo {author} {\bibfnamefont {N.}~\bibnamefont
  {{Bultinck}}}, \bibinfo {author} {\bibfnamefont {M.}~\bibnamefont
  {{Mari{\"e}n}}}, \bibinfo {author} {\bibfnamefont {D.~J.}\ \bibnamefont
  {{Williamson}}}, \bibinfo {author} {\bibfnamefont {M.~B.}\ \bibnamefont
  {{{\c{S}}ahino{\u{g}}lu}}}, \bibinfo {author} {\bibfnamefont
  {J.}~\bibnamefont {{Haegeman}}}, \ and\ \bibinfo {author} {\bibfnamefont
  {F.}~\bibnamefont {{Verstraete}}},\ }\href {\doibase
  10.1016/j.aop.2017.01.004} {\bibfield  {journal} {\bibinfo  {journal} {Annals
  of Physics}\ }\textbf {\bibinfo {volume} {378}},\ \bibinfo {pages} {183}
  (\bibinfo {year} {2017})}\BibitemShut {NoStop}%
\bibitem [{\citenamefont {Sanz}\ \emph {et~al.}(2009)\citenamefont {Sanz},
  \citenamefont {Wolf}, \citenamefont {P{\'e}rez-Garc{\'i}a},\ and\
  \citenamefont {Cirac}}]{Sanz:PhysRevA.79.042308}%
  \BibitemOpen
  \bibfield  {author} {\bibinfo {author} {\bibfnamefont {M.}~\bibnamefont
  {Sanz}}, \bibinfo {author} {\bibfnamefont {M.~M.}\ \bibnamefont {Wolf}},
  \bibinfo {author} {\bibfnamefont {D.}~\bibnamefont {P{\'e}rez-Garc{\'i}a}}, \
  and\ \bibinfo {author} {\bibfnamefont {J.~I.}\ \bibnamefont {Cirac}},\ }\href
  {\doibase 10.1103/PhysRevA.79.042308} {\bibfield  {journal} {\bibinfo
  {journal} {Phys. Rev.}\ }\textbf {\bibinfo {volume} {A79}},\ \bibinfo {pages}
  {042308} (\bibinfo {year} {2009})},\ \Eprint {http://arxiv.org/abs/0901.2223}
  {arXiv:0901.2223 [cond-mat.str-el]} \BibitemShut {NoStop}%
\end{thebibliography}

\def\cprime{$'$}

\appendix
\section{\label{ap:UqSU2}The quantum group $\qsu$}

  There exist different conventions for the quantum group $\qsu$ and hence it is useful to be explicit about the convention we use. Our definition follows Ref.~\onlinecite{Klimyk:MR1492989} even though it should be noted that
  what we call $\qsu$ is called $\breve{\cU}_q(sl_2)$ in that book.

  The quantum group $\qsu$ carries the structure of a Hopf algebra and defines a $q$-deformation of the Lie algebra $su(2)$ which is recovered as $q\to1$. $\qsu$ has an algebra structure that is encoded in the commutation relations~\eqref{eq:CommRels} and a compatible coalgebra structure that is defined in Eq.~\eqref{eq:Coproduct} in terms of a coproduct $\Delta:\qsu\to\qsu\otimes\qsu$. We note that the first two relations of Eq.~\eqref{eq:CommRels} can be written as
\begin{align}
q^{\alpha\bS^z}\,\bS^\pm\,q^{-\alpha\bS^z}
=q^{\pm\alpha}\,\bS^\pm\;.
\end{align}
  From a practical perspective the coproduct permits to define the notion of tensor product representations. To complete the characterization of $\qsu$ we also need to define the unit $\eta:\Complex\to\qsu$, the counit $\epsilon:\qsu\to\Complex$ and the antipode $S:\qsu\to\qsu$. The latter is given by
\begin{align}
  S(\bS^z)
  =-\bS^z\ ,\quad
  S(\bS^\pm)
  =-q^{\pm1}\,\bS^\pm\;.
\end{align}
The other functions have the form $\epsilon\equiv0$ (on the generators $\bS^z$ and $\bS^\pm$) and $\eta(1)=\bI$.

  Since quantum spin chains are defined on a Hilbert space we also need to introduce a suitable notion of hermitian conjugation. This is captured by a Hopf-$\ast$ structure. With our choice of coproduct and for real values of $q$ the Hopf-$\ast$ structure is given by
\begin{align}
\label{eq:Adjoint}
(\bS^z)^\ast
&=\bS^z\;,&
(\bS^\pm)^\ast
&=\bS^\mp\;.
\end{align}
  We note that a different choice of coproduct (as is sometimes preferred in the mathematics literature) would lead to different and physically rather unnatural expressions for the hermitean conjugation.

\section{\label{ap:UqSUreps}Representations of $\qsu$}

  The representation theory of $\qsu$ for real values of $q\neq0$ very much mimics the well-known representation theory of $su(2)$.\cite{Klimyk:MR1492989} All finite dimensional representations are fully reducible. The quantum group has irreducible representations $\cV_j$ that are labelled by a spin $j=0,\frac{1}{2},1,\ldots$ and have dimension $2j+1$. Moreover, the decomposition of a tensor product $j_1\otimes j_2$ precisely corresponds to the well-known decomposition for $su(2)$. In view of the non-trivial (and $q$-dependent) coproduct, concrete expressions for Clebsch-Gordan coefficients are different though.
  
  The representation $\cV_j$ is spanned by (orthonormal) vectors $|m\rangle$ with $m=-j,\ldots,j$ on which the generators of $\qsu$ act by
\begin{align}
  \label{eq:Generators}
  \bS^z|m\rangle
  &=m|m\rangle\nonumber\\
  \bS^+|m\rangle
  &=\sqrt{[j-m]_q[j+m+1]_q}|m{+}1\rangle\\
  \bS^-|m\rangle
  &=\sqrt{[j+m]_q[j-m+1]_q}|m{-}1\rangle\;.\nonumber
\end{align}
  These expressions make use of so-called $q$-numbers $[x]_q$ that are defined via
\begin{align}
  \label{eq:qNumber}
  [x]_q
  =\frac{q^x-q^{-x}}{q-q^{-1}}
  \xrightarrow{\ q\to1\ }x\;.
\end{align}
  An important invariant of the representation $\cV_j$ is the quantum dimension $\qdim(j)$ that is defined by
\begin{align}
  \label{eq:QuantumDimension}
  \qdim(j)
  =\tr_j\bigl(q^{2\bS^z}\bigr)
  =[2j+1]_q
\end{align}
  and approaches the standard dimension $\dim(j)=2j+1$ in the limit $q\to1$. In the context of the \qAKLT model one naturally encounters the representations $\cV_0$, $\cV_{\frac{1}{2}}$ and $\cV_1$ which describe singlet bonds, auxiliary spins and physical spins, respectively.

\section{\label{ap:MPS}Construction of the MPS}

  In the main text we used an expression for the MPS that is different from others that can be found in the literature. We therefore include the simple derivation here. For the construction of the MPS tensor we look at one physical site which is comprised of two auxiliary spins plus the left auxiliary spin of its neighbor to the right. We thus consider the tensor product $\cV_{\frac{1}{2}}\otimes\cV_{\frac{1}{2}}\otimes\cV_{\frac{1}{2}}$ and start by inducing a singlet in the right two factors. This is done by means of the map $I_0:\cV_0\to\cV_{\frac{1}{2}}\otimes\cV_{\frac{1}{2}}$
\begin{align}
I_0(1)
=\frac{1}{\sqrt{q+q^{-1}}}\bigl(q^{\frac{1}{2}}\,|\!\!\uparrow\downarrow\rangle
-q^{-\frac{1}{2}}\,|\!\!\downarrow\uparrow\rangle\bigr)\;.
\end{align}
  Up to normalization this yields the two states
\begin{align*}
  |\alpha\rangle\bigl(q^{\frac{1}{2}}\,|\!\!\uparrow\downarrow\rangle
  -q^{-\frac{1}{2}}\,|\!\!\downarrow\uparrow\rangle\bigr)
  =q^{\frac{1}{2}}
   |\alpha\!\uparrow\rangle|\!\!\downarrow\rangle
   -q^{-\frac{1}{2}}
   |\alpha\!\downarrow\rangle|\!\!\uparrow\rangle\;,
\end{align*}
  where $\alpha\in\{\uparrow,\downarrow\}$. On these states we act with the projector onto the $S=1$ component in the first two factors which can easily be confirmed to be given by
\begin{align*}
  P_1
  =|+\rangle\langle\uparrow\uparrow\!\!|
    +|-\rangle\langle\downarrow\downarrow\!\!|
   +\frac{q^{-\frac{1}{2}}|0\rangle\langle\uparrow\downarrow\!\!|+q^{\frac{1}{2}}|0\rangle\langle\downarrow\uparrow\!\!|}{\sqrt{q+q^{-1}}}
   \;.
\end{align*}
  Writing the resulting states in matrix form in the standard basis of $\cV_{\frac{1}{2}}$ we find
\begin{align}
  g=\mat-\frac{q^{-1}}{q+q^{-1}}\,|0\rangle&\frac{q^{\frac{1}{2}}}{\sqrt{q+q^{-1}}}\,|+\rangle\\
      -\frac{q^{-\frac{1}{2}}}{\sqrt{q+q^{-1}}}\,|-\rangle&\frac{q}{q+q^{-1}}\,|0\rangle\tam\;.
\end{align}
  This expression turns out to be right canonical but not normalized. Including the correct normalization leads to the MPS tensor~\eqref{eq:MPStensorB} used in the main text.

\section{\label{ap:Equivariance}Equivariance of the MPS tensor}
	
  The MPS tensor defined in Eq.~\eqref{eq:MPStensorB} satisfies the equivariance relations
\begin{align}
	\label{eq:BEquivariance}
	\bS^z\triangleright B
	&=\bS^zB-B\,\bS^z\nonumber\\
	q^{\alpha\bS^z}\triangleright B
	&=q^{\alpha\bS^z}Bq^{-\alpha\bS^z}\\
	\bS^\pm\triangleright B
	&=\bS^\mp Bq^{-\bS^z}-q^{\mp1}\,q^{-\bS^z}B\,\bS^\mp\nonumber
\end{align}
  as can easily be checked case by case. We note that these relations can be expressed as $X\triangleright B=B(\idop\otimes S)\Delta(X^\ast)$ (with a suitable interpretation of the action on $B$ on the right hand side) as expected from the general Hopf algebra structure of $\cU_q(sl_2)$. The first two lines reflect what one would have for the action of an ordinary Lie algebra and a group, respectively. (See Ref.~\onlinecite{Sanz:PhysRevA.79.042308} for a discussion of equivariance properties of general MPS tensors in the group case.)
	
  The identities above are valid for a single site. In an MPS one has mixed tensor/matrix products of the form
\begin{align}
	|\text{MPS}\rangle
	=B_1B_2\cdots B_L\;,
\end{align}
  where the index is referring to the site of the physical spin. This product has the equivariance properties
\begin{align}
  \label{eq:FullMPSEquivariance}
  \bS^z\triangleright(B_1\cdots B_L)
  &=\bS^zB_1\cdots B_L-B_1\cdots B_L\bS^z\nonumber\\
      q^{\alpha\bS^z}\triangleright(B_1\cdots B_L)
  &=q^{\alpha\bS^z}B_1\cdots B_Lq^{-\alpha\bS^z}\\
      \bS^\pm\triangleright(B_1\cdots B_L)
  &=\bS^\mp B_1\cdots B_Lq^{-\bS^z}\nonumber\\
  &\qquad-q^{\mp1}\,q^{-\bS^z}B_1\cdots B_L\bS^\mp\;,\nonumber
\end{align}
  which are an immediate consequence of the relations~\eqref{eq:BEquivariance} for the individual tensors. Indeed, the terms created between two $B$s simply drop out. When verifying these relations it is important to work with the correct coproduct for a multiple tensor product.
	
\section{\label{ap:PBC}Periodic boundary conditions}
	
  An important consequence of the relations~\eqref{eq:FullMPSEquivariance} is that an MPS with periodic boundary conditions is not invariant under the action of $\qsu$. Instead one will need to work with the quantum trace
\begin{align}
	|\text{MPS}\rangle
	=\tr\Bigl[q^{2\bS^z}B_1\cdots B_L\Bigr]\;.
\end{align}
  It is evident that this state is invariant under the action of $\bS^z$ and $q^{\alpha\bS^z}$. When acting with $\bS^\pm$ we find
\begin{align}
  \bS^\pm|\text{MPS}\rangle
  &=\tr\Bigl[q^{2\bS^z}\bS^\mp Xq^{-\bS^z}\Bigr]
      -q^{\mp1}\tr\Bigl[q^{2\bS^z}q^{-\bS^z}X\bS^\mp\Bigr]\nonumber\\
  &=\tr\Bigl[q^{\bS^z}\bS^\mp q^{-\bS^z}q^{\bS^z}X\Bigr]
      -q^{\mp1}\tr\Bigl[\bS^\mp q^{\bS^z}X\Bigr]\nonumber\\
  &=q^{\mp1}\tr\Bigl[\bS^\mp q^{\bS^z}X\Bigr]
      -q^{\mp1}\tr\Bigl[\bS^\mp q^{\bS^z}X\Bigr]\nonumber\\
  &=0\;,
\end{align}
  where we have introduced the abbreviation $X=B_1\cdots B_L$. The necessity for the use of quantum traces in a quantum group context has been known for a long time in the context of quantum integrable systems, see e.g.\ Refs.~\onlinecite{Links:1999JMP....40..726L,Couvreur:2017PhRvL.119d0601C}.


\end{document}